\documentclass[twocolumn,showpacs,preprintnumbers,amsmath,amssymb,prl,aps]{revtex4}
\usepackage{graphicx}
\usepackage{natbib}

\usepackage{dcolumn}
\usepackage{amsmath}

\makeatletter
\def\btt#1{\texttt{\@backslashchar#1}}
\DeclareRobustCommand\bblash{\btt{\@backslashchar}}
\makeatother

\usepackage{color}

\begin{document}

\preprint{HEP/123-qed}


\title[Short Title]{Circularly polarized vacuum field in three-dimensional chiral photonic crystals\\
probed by quantum dot emission}

\author{S. Takahashi$^{1, 2}$}\email{shuntaka@kit.ac.jp}
\author{Y. Ota$^1$}
\author{T. Tajiri$^3$}
\author{J. Tatebayashi$^1$}
\author{S. Iwamoto$^{1, 3}$}
\author{Y. Arakawa$^{1, 3}$}

\affiliation{
$^1$Institute for Nano Quantum Information Electronics, University of Tokyo,\\
4-6-1 Komaba, Meguro-ku, Tokyo 153-8505, Japan\\
$^2$Kyoto Institute of Technology, Matsugasaki, Sakyo-ku, Kyoto 606-8585, Japan\\
$^3$Institute of Industrial Science, University of Tokyo, 4-6-1 Komaba, Meguro-ku, Tokyo 153-8505, Japan
}


\date{\today}


\begin{abstract}
The quantum nature of light-matter interactions in a circularly polarized vacuum field was probed by spontaneous emission from quantum dots in three-dimensional chiral photonic crystals.
Due to the circularly polarized eigenmodes along the helical axis in the GaAs-based mirror-asymmetric structures we studied, we observed highly circularly polarized emission from the quantum dots.
Both spectroscopic and time-resolved measurements confirmed that the obtained circularly polarized light was influenced by a large difference in the photonic density of states between the orthogonal components of the circular polarization in the vacuum field.
\end{abstract}



\maketitle


Circular polarization (CP) of light has attracted much attention for applications such as three-dimensional (3D) displays, bio-chemical sensing for chiral molecules, spintronics in solid states, and quantum information technology.
One of the conventional ways of obtaining circularly polarized light, using a linear polarizer and a quarter-wave plate with a linearly polarized light source, induces some loss of energy, which prevents the use of this approach for achieving dense integration of optical circuits.
Therefore, circularly polarized spontaneous emitters on the micrometer-scale have been studied by controlling spin states in solids with external magnetic fields \cite{Ando,Iba}, by using the valley degree of freedom in monolayer chalcogenides \cite{Iwasa}, by using resonant multipolar moments in plasmonic nanoantennas \cite{Kruk}, by modifying the vacuum field with periodic nano-structures \cite{Konishi,Sven}, and by selecting the local density of state (DOS) of the vacuum field in nano-structures \cite{Luxmoore,Petersen,Sollner,Young,Makhonin}.
In particular, control of the vacuum field enables us to tune spontaneous emission, with respect not only to its polarization but also the emission rates and directions \cite{book:Scully}.
However, these studied nano-structures for controlling the vacuum are 1D or (pseudo-)2D structures.
Since the electric/magnetic field in circularly polarized light has 3D helical symmetry, CP can be eigenpolarization over a broad spatial/spectral range only when the geometric structure has 3D helical symmetry \cite{Gansel}.
Thus, additional spatial dimensions in 3D helical/chiral nano-structures will allow sophisticated control of circularly polarized light as well as deep insights into the circularly polarized vacuum field.


In this work, we successfully modified each circularly polarized component in the vacuum field independently by using semiconductor-based 3D chiral photonic crystals (PhCs) \cite{Takahashi1,Takahashi2}, and we demonstrated CP emission from quantum dots (QDs) embedded in the structures.
Both the degree of circular polarization (DOP) of the emitted light and the radiative lifetime for each CP emission were measured to confirm the modulation of CP components in the vacuum field.
Based on the circularly polarized band structure formed by the 3D chiral PhCs, the former measurement showed a DOP as high as $\sim$ 50 \% for each CP at different wavelengths.
The latter measurement at a particular wavelength showed a $\sim$ 10 \% difference in the radiative lifetime between the orthogonal components of the CP emission.
The wavelength dependencies in these two measurements were consistent each other.
Although similar results have been reported in liquid crystals \cite{Woon,Coles}, one of the important pieces of evidence presented for different emission rates between the orthogonal components of the CP has not been investigated in these reports.
In addition, semiconductor-based PhCs can be monolithically compatible with current electrical circuits and future optical circuits, and can be developed into active devices driven by electrical current.
Furthermore, manipulation of electron/hole spins in semiconductor-based nanostructures can be transferred to circularly polarized light through spin-photon interfaces \cite{Greve,Gao}.
Our results obtained in the absence of a magnetic field will pave the way to realize circularly polarized micro-lasers without any spin injection in semiconductor systems, or optical readout system that can read superposed spin states by measuring the emission rates of the CPs.


\begin{figure}[t]
\includegraphics[width=0.9\linewidth]{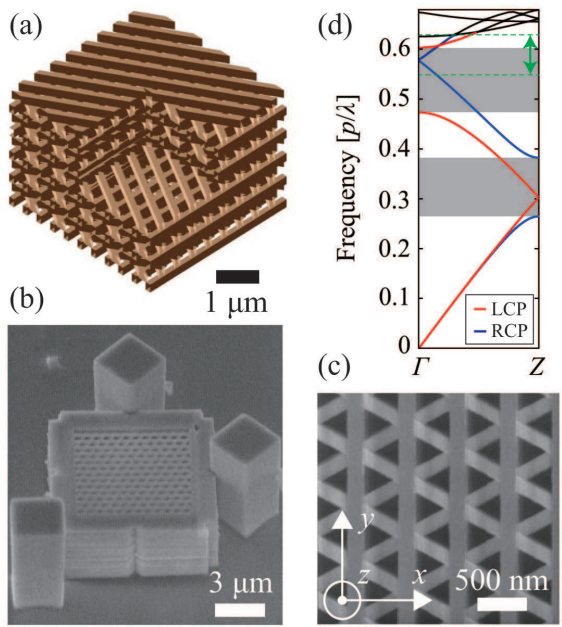}
\caption{\label{fig:SEM}
(a) Schematic diagram of a rotationally-stacked woodpile structure.
The top plate is removed and a part of the structure is cut away for clarity.
(b) Scanning electron micrograph (SEM) image of the fabricated 3D PhC.
Three posts are used as a guide for the micro-manipulation technique.
(c) SEM image of the periodic rods.
The crossing points arranged in a triangular lattice are aligned along the helical axis.
(d) Numerically calculated photonic band structure along the helical axis for the chiral PhC.
The photonic band strongly polarized in LCP (RCP) is colored in red (blue).
The polarization bandgaps for the opposite sense of CP appear in the two shaded regions.
The frequency region indicated by the green double-headed arrow is measured in this study.
}
\end{figure}


One of the studied 3D chiral PhCs composed of a rotationally stacked woodpile structure \cite{Takahashi1,Takahashi2} is shown in Fig. 1(a).
The structure is a stack of patterned thin plates of 225 nm thickness.
These plates made of GaAs were fabricated by electron-beam lithography and dry and wet etching.
The pattern in the plates was an array of rods with 120 nm width and 500 nm period.
These 16 plates were stacked using a micro-manipulation technique \cite{Aoki1,Aoki2,Aniwat} one-by-one with a $60^\circ$ in-plane rotation of each plate in order that three plates construct a single helical unit as a space group with 3$_{1}$ screw operation.
Hence, the pitch of the helix was $p$ = 675 nm.
In this study, we prepared two kinds of plates: active (passive) plates with (without) three layers of InAs self-assembled QDs having a density of $10^{10}$ /cm$^{2}$.
The studied chiral structures contained QDs in the middle three plates.
One of the fabricated chiral PhCs is shown in Fig. 1(b) and (c).
The $x$ ($y$) axis is defined to be orthogonal (parallel) to the rods in the top plate, and the $z$ axis is parallel to the helical axis.
For comparison, we also fabricated an achiral structure by stacking the plates without any in-plane rotation, meaning that the rods were aligned throughout the stacking direction.
Note that the chirality of similar structures was previously confirmed by measuring optical activity, namely, optical rotation \cite{Takahashi1} and circular dichroism \cite{Takahashi2}.
Also, the QDs were naturally $p$-doped due to carbon incorporation during their growth by metal-organic chemical vaper deposition.


The periodically stacked rods formed a photonic band structure along the helical axis, as confirmed in numerical simulations by a plane wave expansion method shown in Fig. 1(d).
Here, $\Gamma$ and $Z$ respectively represent the origin and edge of the first Brillouin zone, corresponding to the $z$ direction in the real space.
Since the optical modes in the red/blue colored bands in Fig. 1(d) were strongly polarized in left/right-handed circular polarization (LCP/RCP), the chiral PhC inherently had two polarization bandgaps in different wavelength regions shaded in Fig. 1(d) \cite{Takahashi2,Lee-Chan}.
In what follows, we focus on one of the polarization bandgaps in the region of 0.48 $<$ $p/\lambda$ $<$ 0.60 (1122 nm $<$ $\lambda$ $<$ 1421 nm) and its band edge at $p/\lambda$ $=$ 0.60 ($\lambda$ $=$ 1122 nm), which is covered by the emission spectrum of the ensemble QDs \cite{EPAPS-Takahashi}.


\begin{figure}[t]
\includegraphics[width=1.0\linewidth]{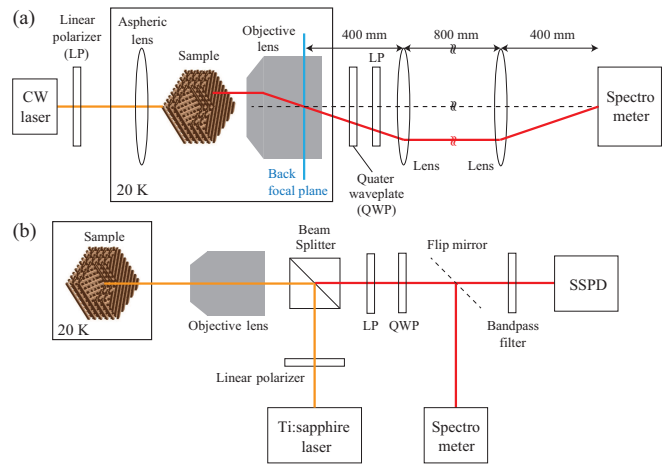}
\caption{\label{fig:Setup}
Schematics diagrams of the experimental setups for the DOP measurement (a) and the radiation lifetime measurement (b).
In (a), the back focal plane is transferred by the 4-f system, and its center part is selected by choosing a single photodiode chip out of the 2D array.
In (b), after checking the emission spectra with the spectrometer, the PL signal at a particular wavelength is selected by a bandpass filter, and its decay curve is measured by the SSPD.
}
\end{figure}


\begin{figure}[t]
\includegraphics[width=1.0\linewidth]{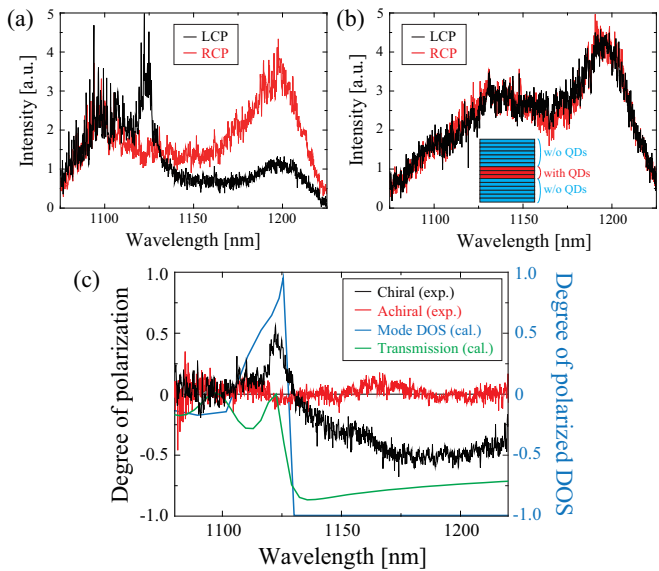}
\caption{\label{fig:DOP}
(a) PL spectra for LCP (RCP) component in black (red) for the chiral PhC containing QDs in the middle three plates.
Clear intensity differences appear at around 1120 nm and 1200 nm in the opposite rotation senses.
(b) PL spectra for LCP (RCP) component in black (red) for the achiral structure.
No intensity difference appears throughout the QD emission wavelength.
Inset shows a schematic diagram of the layered achiral/chiral PhC with QDs in the middle three plates.
(c) DOP as a function of wavelength obtained from (a) in black, from (b) in red, from the $\rho_{LCP/RCP}(\omega)$ in blue, and from the calculated transmittance in green.
At the polarization band edge at around 1120 nm, the QD emission from the chiral structure and the $\rho_{LCP/RCP}$ showed positive DOP values, while the transmission showed zero DOP.
}
\end{figure}


First, we performed photoluminescence (PL) measurements by optically exciting InAs wetting layers near the QDs in the PhCs and by detecting the light emitted from the QDs.
The measurement setup is shown in Fig. 2(a).
The excitation laser light had a power of 40 $\mu$W and the linearly polarized output light was focused on the samples into a 3 $\mu$m spot size by a 60$\times$ aspheric lens having a numerical aperture of 0.54.
The wavelength of the laser light was 980 nm, at which GaAs is transparent.
Then, the broadband infrared luminescence from the backside of the samples was collimated by a 50$\times$ objective lens with a numerical aperture of 0.55.
The collimated luminescence was passed through a quarter-wave plate and a polarizer and was detected by a spectrometer equipped with a 2D array of InGaAs photodiodes.
The samples, as well as the aspheric lens and the objective lens, were cooled down to 20 K, and no magnetic field was applied.
Note that since the circularly polarized band structure was created only in the direction along the helical axis, we selectively detected the emitted light possessing momentum parallel to the helical axis using a 4-f system.
Two additional lenses with a focal length of 400 mm and a numerical aperture of 0.03 were focused on the back focal plane of the objective lens and on the slit of the spectrometer, respectively.
The center region of the back focal plane transferred by this 4-f system was selectively detected by choosing a single InGaAs photodiode of size 20 $\mu$m $\times$ 20 $\mu$m out of the 640 $\times$ 512 2D array.


In order to observe the direct effect of the modulated circularly polarized vacuum field on the QD emission, we also performed time-resolved PL measurements.
The setup is schematically shown in Fig. 2(b).
The QDs in the PhCs were excited by a Ti:sapphire mode-locked laser with a central wavelength of 890 nm, a pulse duration time of 1 ps, a repetition rate of 80 MHz, and a time-averaged power of 1.5 $\mu$W.
The pulsed laser light was linearly polarized and focused in a 3 $\mu$m-diameter spot on the sample by a 50$\times$ objective lens with a numerical aperture of 0.65.
After selecting the wavelength and polarization of the emitted light, a superconductive single photon detector (SSPD) was used for measuring the PL intensity decay curve with a timing resolution of $\sim$ 25 ps.


Figure 3(a) shows the PL intensity of LCP (RCP) in black (red) as a function of wavelength for the chiral PhC containing QDs in the middle three plates.
The LCP intensity was stronger than the RCP intensity in a narrow wavelength region at around 1120 nm, whereas the LCP intensity was weaker than the RCP intensity in a broad wavelength region at around 1200 nm.
Such a clear difference of PL intensity between LCP and RCP did not appear for the achiral structure, as shown in Fig. 3(b).
Note that the two broad peaks at around 1100 nm and 1200 nm were the ensemble QD emission from the first excited state and the ground state, respectively \cite{EPAPS-Takahashi}.


From these results, we calculated the DOP from the detected intensity of LCP and RCP using $DOP = (I_{LCP} - I_{RCP}) / (I_{LCP} + I_{RCP})$.
Figure 3(c) plots the DOP as a function of the emitted wavelength for the chiral (black) and the achiral (red) structures.
This figure clearly shows that the QDs in the chiral PhC emitted narrowband LCP light in the short wavelength region, whereas broadband RCP light was obtained in the long wavelength region.
The absolute value of the obtained DOP was as high as $\sim$ 50 \%, which is sufficiently larger than that for the achiral structure.
Here, we define "mode DOS" $\rho_{LCP/RCP}(\omega)$ of the vacuum field for the particular LCP/RCP electromagnetic mode in Fig. 1(d) at an angular frequency $\omega$.
Note that $\rho_{LCP/RCP}(\omega)$ is only a part of the total DOS, $\rho_{total}(\omega)$, which also includes the DOS of the other modes propagating in the other 3D directions.
In the polarization band gap formed by the chiral PhC, $\rho_{LCP}(\omega)$ was suppressed.
Therefore, the QDs prefer to emit RCP rather than LCP light in the long wavelength region.
On the other hand, in the short wavelength region at around the polarization band edge, $\rho_{LCP}(\omega)$ was greatly enhanced, and LCP light was strongly emitted from the QDs.
In fact, the degree of polarized DOS $(\rho_{LCP} - \rho_{RCP}) / (\rho_{LCP} + \rho_{RCP})$ derived from the band structure in Fig. 1(d) showed positive values at around the band edge, as depicted by the blue curve in Fig. 3(c).


Here, we consider the effect of circular dichroism, which works as a filter for CP in the transmission.
We performed numerical simulations based on the finite-difference time-domain (FDTD) method for the chiral PhC with periodic boundary conditions in the in-plane ($x$ and $y$) directions and with perfectly matched layers attached at 10 $\mu$m away from the center of the structure in the stacking ($z$) direction.
The PhC was normally irradiated with a pulsed plane wave having each CP.
The pulse duration was a single cycle for the central wavelength of 1500 nm.
The recorded time-dependent electromagnetic field of the transmitted light for each CP incidence was analyzed in frequency space using a Fourier transform.
From the obtained transmittance spectrum for each incident CP, the DOP spectrum was plotted (green curve in Fig. 3(c)).
While the QD emission in the chiral PhC showed a positive DOP at the band edge at around 1120 nm, the transmission showed zero DOP.
Even if $\rho_{LCP}$ is enough large, the LCP transmittance reaches unity at most, the same as the RCP transmittance.
Therefore, the sign difference of the measured DOP indicates that the measured CP emission at around 1120 nm resulted from the large $\rho_{LCP}$ at the polarization band edge, rather than the circular dichroism effect.


We also measured other chiral PhCs with different structural parameters, such as different rod widths from 120 nm to 150 nm and 190 nm, and different plate thicknesses from 225 nm to 240 nm.
Since the vacuum field is modified by the CP band structure, tuning the structural parameters of the chiral PhC modulated the band structure, resulting in a shift of the wavelength for the DOP peak/dip in Fig. 3(c).
In fact, the obtained DOP spectra for the chiral PhCs with different structural parameters showed red shifts of the wavelength for the DOP peak and dip \cite{EPAPS-Takahashi}.
These results confirmed that the DOP spectrum in Fig. 3(c) was significantly influenced by the modified circularly polarized vacuum field in the chiral PhC.


\begin{figure}[t]
\includegraphics[width=1.0\linewidth]{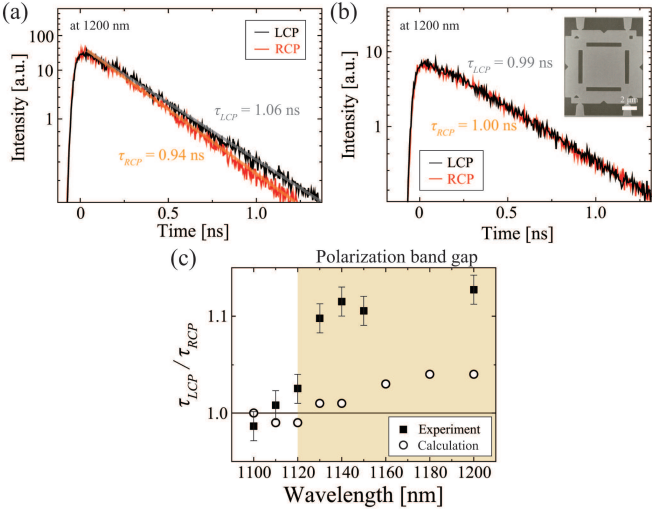}
\caption{\label{fig:Lifetime}
(a) Time-decay of the PL intensity for the two orthogonal components of the CP at 1200 nm wavelength for the chiral PhC.
The PL intensity is normalized for clarity.
The gray and orange lines are single exponential fitted curves for the LCP and RCP decay curves, respectively.
The suppression of the $\rho_{LCP}$ caused an imbalance between LCP and RCP components in $\rho_{total}$, resulting in the longer radiative lifetime for LCP than that for RCP.
(b) Time-decay of the PL intensity at 1200 nm wavelength for a single active plate containing QDs, as shown in the inset.
No difference was observed between the two CPs.
(c) Experimentally obtained radiative lifetime ratio $\tau_{LCP}/\tau_{RCP}$ as a function of wavelength for the chiral PhC (solid squares).
Numerically obtained lifetime ratio is also plotted as open circles, showing qualitative agreement with the experimental results.
A lifetime difference between the CPs appeared in the long wavelength region, which is consistent with the polarization bandgap.
}
\end{figure}


Figure 4(a) shows the time-decay of the PL intensity obtained by the time-resolved PL measurements for the chiral PhC at a wavelength of 1200 nm in the polarization bandgap.
By fitting with a single exponential function at around the beginning of the intensity decay, the radiative lifetime of LCP, $\tau_{LCP} = 1.06$ ns was longer than that of RCP, $\tau_{RCP} = 0.94$ ns.
Since radiative lifetime is proportionally influenced by the total DOS $\rho_{total}$ \cite{book:Scully,book:Novotny}, the lifetime difference indicates that $\rho_{LCP}$ is smaller than $\rho_{RCP}$ in the polarization bandgap.
Note that there was no difference between $\tau_{LCP}$ and $\tau_{RCP}$ for an active plate without the rod pattern, as shown in Fig. 4(b), confirming that the lifetime difference of the CPs in Fig. 4(a) originates from the chiral structure.
Figure 4(c) plots the radiative lifetime ratio $\tau_{LCP}/\tau_{RCP}$ for various wavelength in the chiral PhC.
A lifetime difference between LCP and RCP appeared in the wavelength region longer than 1120 nm, which is consistent with the polarization band edge obtained in Fig. 3(c).


We performed numerical calculations of the lifetime ratio \cite{EPAPS-Takahashi}.
Using the FDTD method, the Poynting vectors were integrated over a surface enclosing the circularly polarized light source in the chiral PhC \cite{Lee-Yariv,Xu-Yariv}.
The numerical results plotted in Fig. 4(c) show that $\tau_{LCP}/\tau_{RCP}$ $>$ 1 in the long wavelength region, which is qualitatively consistent with the experimental results.
The quantitative difference between the experimental and numerical results was partly caused by the small in-plane size of the numerical chiral PhC, owing to computational time restrictions \cite{EPAPS-Takahashi}.


In spite of the high DOP $\sim$ 50 \% in Fig. 3(c), the obtained lifetime difference between the orthogonal components of the CP was only $\sim$ 10 \% in both the experiment and the calculation.
This is because the DOP is influenced only by $\rho_{LCP/RCP}(\omega)$ for particular modes propagating in the $z$ direction due to the selection of momentum using a single photodiode.
On the other hand, the lifetime is derived from the total DOS $\rho_{total}$ which is the integral for all modes propagating in any 3D direction and having any polarization.
Hence, the large difference between $\rho_{LCP}$ and $\rho_{RCP}$ was almost averaged by the DOS for all other modes.
The subsequent selection of momentum by choosing a single photodiode out of the 2D array did not affect the lifetime.


Another possible reason for the small lifetime difference compared with the DOP is the effect of spin relaxation in the QDs, and this should be taken into account.
Rate equations for the population of electron-hole pairs emitting LCP (RCP), $P_{LCP}$ ($P_{RCP}$), including spin relaxation in the absence of a magnetic field, are written as \cite{Seymour},
\begin{eqnarray}
\frac{dP_{LCP}}{dt} &=& -\frac{P_{LCP}}{\tau_{LCP}}-\frac{P_{LCP}-P_{RCP}}{\tau_{spin}}, \\
\frac{dP_{RCP}}{dt} &=& -\frac{P_{RCP}}{\tau_{RCP}}+\frac{P_{LCP}-P_{RCP}}{\tau_{spin}}.
\end{eqnarray}
Here, $\tau_{spin}$ is the electron/hole spin relaxation time $T_1$.
For both electrons and holes in ensemble self-assembled QDs, $T_1$ $>$ $T_2$ $>$ $T_2^*$ = 1 $-$ 10 ns \cite{Kloeffel} for spin decoherence (dephasing) time $T_2$ ($T_2^*$).
This time scale is comparable to the obtained radiative lifetime for both CPs.
In these conditions, as well as $P_{LCP}$ = $P_{RCP}$ at $t$ = 0 because of the linearly polarized excitation pulse, the rate equations showed a significant difference in time-decay between $P_{LCP}$ and $P_{RCP}$ even in the presence of the spin relaxation.
Note that under the condition that the spin relaxation time is much shorter than $\tau_{LCP(RCP)}$, such as electron-hole exchange interactions in non-doped QDs, the time-decays of $P_{LCP}$ and $P_{RCP}$ coincide.
This was not the case in this study where positively-charged excitons $X^+$ were mainly involved due to the natural $p$-doping in the QDs.
Therefore, we can conclude that the spin relaxation in the QDs has little effect on the measured lifetime difference.
The lifetime difference could be enhanced by band engineering in the in-plane directions, or by employing different structures that have helical axes in various directions, such as the {\bf srs} net \cite{Saba}.


At the polarization band edge where $\rho_{LCP}$ was enhanced, both the time-resolved PL measurement and the calculation showed $\tau_{LCP}/\tau_{RCP}$ $\sim$ 1 in Fig. 4(c), whereas $\tau_{LCP}/\tau_{RCP}$ $>$ 1 in the polarization bandgap where $\rho_{LCP}$ was suppressed.
This difference is due to the spatial distribution of the electromagnetic field.
Another numerical calculation showed that the in-plane distribution of the LCP electric field at the band edge was relatively localized in the rods compared with that at the band gap \cite{EPAPS-Takahashi}.
Therefore, by considering the local DOS in the rods, $\tau_{LCP}/\tau_{RCP}$ at the band edge is sensitive to averaging for all QD positions in the rods and reaches unity, whereas $\tau_{LCP}/\tau_{RCP}$ $>$ 1 is maintained in the band gap.
A similar averaging effect for the lifetime or DOS at band edges has been experimentally observed and discussed in 2D PhCs \cite{Fujita,Wang}.
However, if a single QD could be positioned accurately in the chiral PhC, the band edge effect would potentially show a large reduction of $\tau_{LCP}$.
In fact, $\tau_{LCP}/\tau_{RCP}$ was shown to be as low as 0.83 at the band edge for a particular position of a light source \cite{EPAPS-Takahashi}.


In conclusion, we investigated spontaneous emission from QDs embedded in 3D chiral PhCs, and confirmed the modulation of the CP components in the vacuum field.
The QDs in the chiral structure emitted highly circularly polarized light with a DOP $\sim$ 50 \% in each sense of CP (right- and left-handed) at the polarization band edge and in the polarization bandgap, respectively.
From the results of spectroscopic and time-resolved measurements, the obtained circularly polarized light was revealed to be directly influenced by the modified DOS of the vacuum field.
These results could be a large step towards the realization of circularly polarized light emitting diodes, lasers, and sensors in semiconductor systems.


This work was supported by Grant-in-Aid for Scientific Research (16H06085, 26889018), Grant-in-Aid for Specially Promoted Research (15H05700), and Grant-in-Aid for Scientific Research on Innovative Areas (15H05868).




\begin{thebibliography}{00}

\bibitem{Ando}
H. Ando, T. Sogawa, and H. Gotoh, Appl. Phys. Lett. {\bf 73}, 566 (1998).

\bibitem{Iba}
S. Iba, {\it et al.}, Appl. Phys. Lett. {\bf 98}, 081113 (2011).

\bibitem{Iwasa}
Y. J. Zhang, {\it et al.}, Science {\bf 344}, 725 (2014).

\bibitem{Kruk}
S. S. Kruk, {\it et al.}, ACS Photonics {\bf 1}, 1218 (2014).

\bibitem{Konishi}
K. Konishi, {\it et al.}, Phys. Rev. Lett. {\bf 106}, 057402 (2011).

\bibitem{Sven}
S. V. Lobanov, {\it et al.}, Phys. Rev. B {\bf 92}, 205309 (2015).

\bibitem{Luxmoore}
I. J. Luxmoore, {\it et al.}, Phys. Rev. Lett. {\bf 110}, 037402 (2013).

\bibitem{Petersen}
J. Petersen, J. Volz, and A. Rauschenbeutel, Science {\bf 346}, 67 (2014).

\bibitem{Sollner}
I. S$\rm \ddot{o}$llner, {\it et al.}, Nat. Nanotech. {\bf 10}, 775 (2015).

\bibitem{Young}
A. B. Young, {\it et al.}, Phys. Rev. Lett. {\bf 115}, 153901 (2015).

\bibitem{Makhonin}
R. J. Coles, {\it et al.}, Nat. Commun. {\bf 7}, 11183 (2016).

\bibitem{book:Scully}
M. O. Scully and M. S. Zubairy, {\it Quantum Optics} (Cambridge University Press, Cambridge, 1997).

\bibitem{Gansel} J. K. Gansel, {\it et al.}, Science {\bf 325,} 1513 (2009).

\bibitem{Takahashi1}
S. Takahashi, {\it et al.}, Opt. Express {\bf 21}, 29905 (2013).

\bibitem{Takahashi2}
S. Takahashi, {\it et al.}, Appl. Phys. Lett. {\bf 105}, 051107 (2014).

\bibitem{Woon}
K. L. Woon, {\it et al.}, Phys. Rev. E {\bf 71}, 041706 (2005).

\bibitem{Coles}
H. Coles and S. Morris, Nat. Photon. {\bf 4}, 676 (2010).

\bibitem{Greve}
K. D. Greve, {\it et al.}, Nature {\bf 491}, 421 (2012).

\bibitem{Gao}
W. B. Gao, {\it et al.}, Nature {\bf 491}, 426 (2012).

\bibitem{Aoki1}
K. Aoki, {\it et al.}, Nat. Mater. {\bf 2}, 117 (2003).

\bibitem{Aoki2}
K. Aoki, {\it et al.}, Nat. Photon. {\bf 2}, 11 (2008).

\bibitem{Aniwat}
A. Tandaechanurat, {\it et al.}, Nat. Photon. {\bf 5}, 91 (2011).

\bibitem{Lee-Chan}
J. C. W. Lee and C. T. Chan, Opt. Express {\bf 13}, 8083 (2005).

\bibitem[{EPA()}]{EPAPS-Takahashi}
\bibinfo{note}{See EPAPS Document No.XXXX for characterization details and additional data.}

\bibitem{book:Novotny}
L. Novotny and B. Hecht, {\it Principles of Nano-Optics} (Cambridge University Press, Cambridge, 2012).

\bibitem{Xu-Yariv}
Y. Xu, {\it et al.}, J. Opt. Soc. Am. B {\bf 16}, 465 (1999).

\bibitem{Lee-Yariv}
R. K. Lee, {\it et al.}, J. Opt. Soc. Am. B {\bf 17}, 1438 (2000).

\bibitem{Seymour}
R. J. Seymour and R. R. Alfano, Appl. Phys. Lett. {\bf 37}, 231 (1980).

\bibitem{Kloeffel}
C. Kloeffel and D. Loss, Annu. Rev. Condens. Matter Phys. {\bf 4}, 51 (2013).

\bibitem{Saba}
M. Saba, {\it et al.}, Phys. Rev. Lett. {\bf 106}, 103902 (2011).

\bibitem{Fujita}
M. Fujita, {\it et al.}, Science {\bf 308}, 1296 (2005).

\bibitem{Wang}
Q. Wang, S. Stobbe, and P. Lodahl, Phys. Rev. Lett. {\bf 107}, 167404 (2011).

\end{thebibliography}
\end{document}